# Resolving chemical structures in scanning tunnelling microscopy

C. Weiss[1,2], C. Wagner[1,2], C. Kleimann[1,2], F.S. Tautz[1,2], R. Temirov[1,2]

[1] *Institut für Bio- und Nanosysteme 3, Forschungszentrum Jülich, 52425 Jülich, Germany*

[2] *JARA – Fundamentals of Future Information Technology*

**With the invention of scanning probe techniques, direct imaging of single atoms and molecules became possible [1]. Today, scanning tunnelling microscopy (STM) routinely provides angstrom-scale image resolution [2-6]. At the same time, however, STM images suffer from a serious drawback - the absence of chemical information. Here we demonstrate a modification of STM that resolves the chemical structure of molecular adsorbates (Fig. 1). The key to the new STM mode is a combined force sensor and signal transducer that is formed within the tunnelling junction when a suitable gas condenses there. The method probes the repulsive branch of the surface adsorption potential and transforms the force signal into a current. Obtained images achieve the same resolution as state-of-the-art atomic force microscopy (AFM) [7]. In contrast to AFM, however, our (uncalibrated) force sensor is of nanoscale dimensions and therefore intrinsically insensitive to those long-range interactions that make atomic-resolution AFM so demanding [8-11].**

The force imaging mode is achieved by condensing molecular hydrogen ($H_2$) or deuterium ($D_2$) in the cold (T <10 K) junction of an ultra-high vacuum STM and is therefore called scanning tunnelling hydrogen microscopy (STHM) [12]. $H_2$ ($D_2$) adsorption abruptly changes the STM image contrast (cf. Fig. 2a). At low coverages the imaging tends to change back spontaneously to the conventional mode after a short time. However, as the $H_2$ ($D_2$) coverage increases, the STHM mode becomes stable for arbitrarily long times. Operating the STM in the new regime, we have imaged prototypical organic molecules adsorbed on noble metal surfaces, in particular Ag and Au (Fig. 1a-c), at constant tip height. The obtained contrast closely corresponds to the chemical structure of the molecule in question, i.e. to its atomic-scale geometry. Fig. 1d shows that STHM can even image intermolecular bonds. Remarkably, the imaging in STHM is *not* sensitive to the local density of states (LDOS) that is conventionally imaged by STM.

The spectroscopic characterisation of the STM junction in the presence of the adsorbed $H_2$ ($D_2$) is the key to understanding STHM. Therefore, we will begin by analysing the junction's differential conductance as a function of bias voltage and tip-sample distance. In conjunction with the structure of the STHM junction that will be deduced from the adsorption properties of molecular hydrogen in a second step, the spectroscopic analysis finally allows us to understand the mechanism of force imaging in STHM.

We start by analysing the question whether there are distinct spectral features in the differential conductance (dI/dV) that are connected to the geometric STHM contrast. We have recorded dI/dV spectra with a static (i.e. not laterally scanning) tip on a grid of 64 x 64 pixels above a single PTCDA molecule on Au(111). Two typical spectra are shown in Fig. 2f. They reveal strong non-linearities in I(V) that are a generic feature of nanoscale junctions in the presence of $H_2$ or $D_2$; their specific shape is a fingerprint of the junction state [12-15]. The dI/dV(x,y,V) data cube allows us to reconstruct dI/dV images at any chosen value of the bias voltage (Fig. 2b-e). Two observations are apparent: Firstly, we see that the geometric STHM contrast arises from lateral variations in the zero-bias differential conductance (Fig. 2b). Secondly, Fig. 2c-e reveal that the geometric STHM contrast disappears when the bias voltage increases beyond a critical value (~40 meV in the present case) that coincides with noisy spikes in dI/dV (Fig. 2f).

The data in Fig. 2 also reveal important insights about the STHM junction itself: The fact that the static image in Fig. 2b is identical to the scanned image in Fig. 1a proves that the STHM imaging mechanism is based on a structurally equilibrated junction; specifically, the geometric STHM contrast is independent of scanning direction and speed, which demonstrates that the processes involved are fast on the time-scale of the measurement. Moreover, Fig. 2f shows that the equilibrium state of the junction, as indicated by the specific shape of its differential conductance spectrum, changes from point to point on the adsorbed molecule.

From the data presented so far the following picture emerges: $H_2$ ($D_2$) adsorption changes the properties of the junction such that its conductance becomes sensitive to the atomic-scale geometry (in contrast to the LDOS) of the surface. When the tip is scanned over the surface, the junction responds to lateral variations of the surface structure by a changing its equilibrium state, thus *sensing* the surface structure; the changed equilibrium state of the junction in turn modifies its (zero-bias) conductance, thereby *transforming* information about the surface structure into a conductance signal which is easily measured in STM. Our task is to understand how (and exactly what about the structure) the junction senses and how it transforms this information into the conductance channel. In other words, we must identify the sensor/transducer in the junction and understand its mechanism.

To this end, we have characterized the STHM junction above PTCDA/Au(111) with a sequence of dI/dV spectra measured at different tip-surface separations (Fig. 3a). From this set of spectra, approach curves at constant bias can be generated (dots in Fig. 3b). Similarly, we have recorded conductance curves at fixed bias as the tip approaches the surface (lines in Fig. 3b). As expected, these approach curves coincide with the ones derived from Fig. 3a, but they have many more data points. The calibration of the tip-surface distance in Fig. 3 is described in the methods section.

The dI/dV spectra in Fig. 3a again display the I(V) non-linearities already known from Fig. 2f. A glance at Fig. 3b reveals that these non-linearities are paired with a marked deviation from the exponential distance dependence which is characteristic of tunnelling. In contrast, for bias voltages beyond the conductance spikes (the position of which varies between ±40mV and ±100mV in Fig. 3a, depending on tip-sample distance) the spectra are featureless and the simple exponential distance dependence is recovered. We note that this recovery coincides with the loss of geometric STHM contrast (Fig. 2d-e). The conductance spikes therefore appear as critical points at which the junction properties change profoundly. Here it is interesting to note that in the context of hydrogen-containing nanojunctions it has been found that conductance spikes appear at the point when hydrogen molecules are excited by the current from a well-defined bound state in the junction to a dense spectrum of loosely bound states outside [15]. This makes clear why the geometric STHM contrast in our experiments disappears: At the spikes the hydrogen is excited out of the junction, and the empty junction reverts to simple tunnelling and conventional LDOS imaging. The energy of the excitation is defined by the local potential which in turn is related to the microscopic shape of the tip, the local structure of the surface, the tip-surface distance and the hydrogen coverage. Due to this complex dependence the dI/dV(V) spectra measured in the presence of $H_2$ ($D_2$) become intrinsically irreproducible, although their generic features such as spikes are always present [12-15]. In contrast to the spectra the STHM imaging is much more robust.

Because we have seen that the deuterium-induced non-exponential distance dependence of the differential conductance at low bias voltages in Fig. 3b is closely related to the geometric STHM contrast, we analyse it further. Using the approach curves shown in Fig. 3b we can quantify the effect of $D_2$ on the conductance by subtracting the low-bias approach curve (0 mV) from the high-bias one (averaged from 120 to 130 mV); the resultant *excess conductance curve* yields a quantitative measure of the deviation from pure exponential behaviour. Fig. 3c displays the excess conductance for different lateral positions above a PTCDA molecule on Au(111). Each position exhibits a distinct and reproducible excess curve, in which two opposing tendencies are always present, one of them enhancing the excess conductance with decreasing tip-sample distance, the other reducing it in the same direction until it even becomes negative. According to Fig. 3b-c, the first tendency dominates at large tip-sample distances (> 7 Å), the latter at smaller ones (7 to 6.4 Å). The competition between the two tendencies leads to a maximum excess at ≈7 Å and allows us to define three characteristic regimes (Fig. 3b).

The excess conductance curves are a fingerprint of the STHM junction's structural evolution upon tip approach; they will form the basis of our analysis of force sensing and transduction in the junction. However, before we can turn to this analysis we must recall the adsorption properties of $H_2$ and $D_2$ and understand their structure and bonding in the STHM junction.

The adsorption of $H_2$ and $D_2$ on noble metals at low temperatures has been studied in great detail [16]. Both experiment and theory have revealed the adsorption to be purely physisorptive. On flat surfaces, the multilayer desorption temperature of $D_2$ is less than 8K, while for $H_2$ not more than two layers can be condensed at temperatures above 4.8 K [17]. This indicates that the coverage in our experiments is self-limiting itself to very few layers of physisorbed gas.

For tip-surface distances > 10 Å, both tip and sample bind $H_2$ ($D_2$) molecules in their respective physisorption wells. The wells are sketched schematically in Fig. 4a. In contrast, at the tip-surface distances in our experiments (<10 Å) the junction can accommodate at most one monolayer of $H_2$ ($D_2$) molecules (Fig. 4b-d), because the typical adsorption height of $H_2$ is 3.2 Å ref. [18], and the typical hydrogen-hydrogen distance is 3.3 Å (for the condensed bulk [19]). On tip approach, the physisorption wells of tip and sample must therefore merge (Fig. 4b), forming a single cavity which confines a single $H_2$ ($D_2$) molecule. Indeed, the high lateral resolution of the geometric STHM images in Fig. 1 suggests that imaging is performed by a single $H_2$ or $D_2$ molecule in the junction that is confined close to the foremost Au tip atom. This picture agrees well with (i) the enhanced binding activity of gold adatoms towards $D_2$ ref. [20] (ii) the appearance of the conductance spikes, which are only to be expected if the $H_2$ or $D_2$ molecule has a stable, well-defined adsorption structure in the junction [15], and (iii) the robustness and universality of the STHM imaging mode itself.

Based on this knowledge we may conclude that the geometric STHM contrast already appears at relatively low gas coverages, as soon as a single $H_2$ ($D_2$) molecule is trapped within the junction. This is indeed observed (Fig. 2a), but the figure also shows that the molecule can disappear spontaneously from the junction during scanning. Our experiments show that increasing the coverage stabilizes the geometric STHM contrast, either via an increased frequency of trapping events and/or by the additional confining potentials of $H_2$ ($D_2$) molecules in the neighbourhood of the junction. Incidentally, the influence of those neighbouring molecules on the junction can be directly monitored by the evolution of the dI/dV spikes as a function of exposure time, even after the STHM mode has established itself [12].

We conclude that the single $H_2$ or $D_2$ molecule in the combined potential cavity of tip and sample is the nanoscale sensor/transducer that is responsible for the geometric STHM contrast. To understand its functionality we now turn back to the excess conductance curves and discuss the origins of the three regimes mentioned above.

The abrupt and irreproducible conductance changes in regime 3 of Fig. 3b arise because the $H_2$ ($D_2$) molecule is mechanically squeezed out of the junction (Fig. 4e), possibly accompanied by the deformation of the electrodes. Directly preceding its forced escape from the junction, the $H_2$ (D$_2$) molecule is gradually "compressed" between the electrodes (Fig. 4c); this occurs in regime 2. We note that a large compressibility is a generic feature of H$_2$ and D$_2$; it is associated with their pronounced zero point motion (ZPM) [19]. When in regime 3 the molecule escapes from the junction, the energy associated with ZPM has become large enough to overcome the lateral confinement. During the compression in regime 2 (Fig. 4c), the ZPM will rise as a consequence of increased Pauli repulsion between the H$_2$ (D$_2$) molecule and the electrodes. Pauli repulsion between a closed shell molecule and a metallic surface in turn depletes the density of states (DOS) in a narrow window around the Fermi level [21-23] (Fig. 4e). As the repulsive interaction gets stronger, the DOS depletion will become more pronounced, thereby quickly decreasing the junction excess conductance, as observed in Fig. 3b, regime 2.

The increasing excess conductance in regime 1 can also be traced back to the evolution of the sensor's confining potential. Regime 1 begins with large tip-surface distances for which both the tip and the surface each bind a H$_2$ (D$_2$) molecule (Fig. 4a). However, as was discussed above, at distances typical for the tunnelling experiment (<10 Å) ref. [2] the junction can only accommodate a single H$_2$ (D$_2$) molecule (Fig 4b). Thus in regime 1 the confining potential will have a characteristic double-well shape along the vertical axis. In this potential, a single H$_2$ (D$_2$) molecule can jump between two local minima. It is conceivable that this motion couples to the electron current through the junction, thereby increasing its conductance [24]. Alternatively, the increased conductance may be explained by off-resonance electron tunnelling through the H$_2$ (D$_2$) molecule confined in the junction [25].

With the junction structure and the interpretation of its spectroscopic signatures in place, we are in a position to explain the geometric STHM imaging mode. Fig. 3d displays a sequence of STHM images acquired at different tip-sample distances. It clearly shows that the geometric contrast is obtained in regime 2, with the best resolution at distances directly preceding regime 3. The origin of the contrast in this regime is shown schematically in Fig. 4c-d. As the tip moves at constant height from a point with lower electron density (Fig. 4c) to point with higher electron density (Fig. 4d), the Pauli repulsion from the adsorbate (represented by a single benzene ring in Fig. 4) will increase. As a result, the confining potential of the H$_2$ (D$_2$) molecule will become steeper on the side of sample, pushing the H$_2$ (D$_2$) molecule closer to the tip. The H$_2$ (D$_2$) molecule thus acts as a sensor of the repulsive interaction above the adsorbate. If the sensor comes closer to the tip, their mutual Pauli repulsion will increase. As a consequence, the $H_2$ ($D_2$) molecule, now acting as a transducer, polarizes and decreases the tip DOS at the Fermi level, leading to a decreased conductance of the STHM junction. The geometric STHM contrast is thus sensitive to total electron densities and can be used to image the chemical structure of molecules (Fig. 1a-c) and intermolecular bonds (Fig. 1d).

According to our model, the geometric STHM contrast arises as a modulation on top of the normal LDOS contrast: As the tip is moved laterally at constant height, the LDOS will change, leading to a different tunnelling conductance. In Fig. 3b, this would correspond to a shift of the exponential curve. At the same time, the confining potential of the sensor will also change, yielding a different excess conductance. As long as the change of excess conductance is larger than the LDOS-induced change of the background conductance itself, the image will be dominated by the geometric STHM contrast. We note that in most cases studied here, the molecule-induced DOS at the Fermi level is small; correspondingly, the LDOS contrast is expected to be faint (cf. Fig. 2a for PTCDA/Au(111)) and we expect the Pauli contrast to dominate. On the other hand, we also have investigated PTCDA/Ag(111) for which substantial molecule-induced LDOS is found at the Fermi level. In this case, depending on the exact parameters, one may expect to observe a superposition of both type of contrast, which indeed we sometimes do.

We finally stress that the STHM method establishes a new paradigm for STM experiments. In conventional STM at typical (not to small) operating distances, the tip is a passive element used to measure the tunnelling probability between its position and the sample; since the tunnelling probability is proportional to the LDOS, information about the sample can be gained. In the novel mechanism described here, the tunnelling current between the passive metal tip and the sample is still the basis of imaging. However, a compliant element is added to the junction. This element is sensitive to a laterally varying sample property other than the LDOS. In the present example of STHM, the Pauli repulsion is relevant, but in principle any other property could be chosen. Because of its compliance, the sensor element (via some appropriate transduction mechanism) changes the junction conductance. In the present example this happens via local polarisation of the tip DOS, but again other transduction mechanisms are conceivable. We therefore suggest that the STHM-based "Pauli repulsion microscopy" with physisorbing gas molecules discussed here is just one example of a broader class of novel STM methods which wait to be discovered.

Acknowledgement

We gratefully acknowledge helpful discussions with J. Kroha (Universität Bonn), M. Rohlfing (Universität Osnabrück), S. Blügel (Forschungszentrum Jülich), and N. Atodiresei (Forschungszentrum Jülich).

**Fig. 1 Chemical structure of organic adsorbates recorded with geometric STHM contrast.** STHM images of **a** 3,4,9,10-perylenetetracarboxylic-acid-dianhydride (PTCDA) adsorbed on Au(111), **b** Pentacene/Ag(111), **c** Sn-Phthalocyanine/Ag(111) shown with the respective chemical structure formulae, **d** 3D image of a PTCDA layer on Au(111). The molecular backbone and oxygen atoms (red dots) are superimposed. Regions of hydrogen bonding are shaded green. A corresponding 2D image is shown in the supplement (Fig. S1). *Imaging parameters:* **a** 1.3x0.7 nm$^2$, constant height, $V_b$=-5 mV, measured with $D_2$. **b** 1.5x0.6 nm$^2$, const. height, $V_b$=-3 mV, $D_2$. **c** 1.5x1.5 nm$^2$, const. height, $V_b$=-5 mV, measured with $H_2$. **d** 3.2x3.2 nm$^2$, const. height, $V_b$=-10 mV, $D_2$. All images as measured, generated with WSxM [26].

**Fig. 2 Coverage and bias dependence of the geometric STHM contrast. a** Spontaneous switching between the conventional LDOS and geometric STHM contrasts at low coverage of $D_2$. **b-e** 64x64 pixel, 1.3x1.3 nm$^2$ dI/dV constant height STHM images extracted from the spectroscopic data acquired over PTCDA/Au(111) with $D_2$. At each pixel of the image one dI/dV spectrum was recorded, using lock-in detection (modulation amplitude 4 mV, frequency 4.8 kHz, spectrum acquisition time 1 s). Minimum (black) and maximum (white) differential conductances are given in images, in units of the conductance quantum $G_0$=2$e^2$/h. Negative conductance values are caused by sharp conductance spikes (see text). **f** dI/dV spectra measured at the marked locations in b). Right panel: spectra as measured. Left panel: Spectra averaged over red and blue circles (diameter 3 pixels ≈ 1 Å) in b). Sharp noise features beyond 20 mV bias are conductance spikes associated with the excitation of the $D_2$ molecule out of the junction (cf. text).

**Fig. 3 Distance dependence of junction conductance and geometric STHM contrast. a** dI/dV spectra measured at the centre of PTCDA on Au(111) with $D_2$ at different tip-surface distances (step 0.1 Å), recorded with lock-in detection (10 mV modulation, frequency 2.3 kHz). Regions

shaded in light grey highlight the conductance spikes (cf. text and the supplement). **b** Differential conductance measured at the centre of PTCDA on Au(111) with $D_2$ with an approaching tip at fixed biases of -5 mV (magenta line) and 120 mV to 130 mV (black line). The black line was averaged over four measured spectra as ($^{dI}/_{dV}$(-130mV, z)+$^{dI}/_{dV}$(-120mV, z)+ $^{dI}/_{dV}$ (120mV,z)+ $^{dI}/_{dV}$ (130mV,z))/4. Data points (black and magenta dots) have been extracted from regions shaded in black and magenta in a). The excess conductance curve (cf. text) calculated by subtracting the black from the magenta curve is shown in light blue. **c** Excess conductance curves as in b) but approaching the tip at positions marked in the inset in the respective colours. Black triangles mark the tip-sample distances at which the images in d) were recorded. **d** Geometric STHM images of PTCDA/Au(111) measured with $D_2$ at different heights indicated by black triangles in c). Imaging parameters: 1.3x1.3 nm$^2$, Constant height, $V_b$=-5 mV. I/V conductance scales (from black to white): 1) $5 \times 10^{-4} < G/G_0 < 3 \times 10^{-3}$, 2) $3 \times 10^{-5} < G/G_0 < 3 \times 10^{-3}$, 3) $5 \times 10^{-5} < G/G_0 < 5 \times 10^{-3}$, 4) $6 \times 10^{-5} < G/G_0 < 6 \times 10^{-3}$, 5) $7 \times 10^{-5} < G/G_0 < 7 \times 10^{-3}$, 6) $2 \times 10^{-4} < G/G_0 < 2 \times 10^{-2}$.

**Fig. 4 Structure and function of the STHM junction.** The junction consists of the atomically sharp noble metal tip (yellow) and the surface (gray) with an aromatic molecule (black) adsorbed on it. In addition, $H_2$ or $D_2$ molecules (small red circles) are present in the junction which is kept at T= 5-10 K. In c) and d) they act as force sensor/transducer (cf. text). The physisorption wells in which $H_2$ or $D_2$ are confined are schematically shown in blue. The z-dependence of the confining potential is shown schematically on the right of each junction, with ground state levels marked in black. **a** At tip-surface distances >10 Å two molecules may physisorb separately on tip and sample. The sample offers a number of binding sites, one such site being located in the centre of $C_6$ rings [18]. The tip will bind a molecule close to its apex [20]. $A_1$, $A_2$ and B are equilibrium bonding distances which add up to ≈10 Å. In this configuration, the tunnelling current is too low to be detected in our setup. **b** At distances between 10 Å and 7 Å the tip and the surface adsorption potentials merge, forming a well which confines a single molecule. The junction is in the regime of enhanced conductance (regime 1 in Fig. 3b, cf. text). **c** At distances between 7 Å and 6.5 Å the confining volume becomes smaller. Due to the increased Pauli repulsion the electrodes (mainly the tip) become polarized and their density of states (DOS) decreases locally, indicated schematically by red shading in the tip. **d** Compared to c), the tip has moved to a position above PTCDA with a larger electron density, corresponding to the C-C bond. The increased Pauli repulsion is shown in the potential diagram. It leads to a stronger confinement and a larger tip polarisation, thus forming the geometric STHM contrast (cf. text). **e** At tip-surface distances < 6.5 Å the sensor becomes unstable because the $H_2$ ($D_2$) molecule escapes from the junction.

**Methods**

As in any STM experiment, the tip structure is also important in STHM. Usually, clean tips that produce conventional LDOS images of acceptable quality also yield images with good geometric STHM contrast. If the tip does not yield good STHM images, dipping it into the metal (2-3 Å, at bias voltages between -0.1 and 0.1 V) in the presence of $H_2$ or $D_2$ can improve the situation. After a few of such tip preparations, a good geometric STHM contrast usually appears.

The absolute distance scale in Fig. 3 was calibrated by the onset of regime 3 in which the junction experiences abrupt structural changes. From simple geometric considerations such changes are expected when the tip-surface distance becomes less than the sum of binding distances of the confined molecule to the electrodes. We assume the binding distance to be 3.2 Å [18] on both sides of the junction. In this way, the tip-surface distance at the beginning of regime 3 is defined to be 6.4 Å. We note that this simple calibration can only give a rough estimate of the correct distances. However, none of our conclusions requires the accurate knowledge of the correct scale.

We note that the regime of geometric STHM imaging must be distinguished from the imaging of compact monolayers of condensed $H_2$ and $D_2$ on metal surfaces [13]. Since we never observe images of static $H_2$ ($D_2$) layers under the conditions studied here, we conclude that our experiments always operate at sub-monolayer coverages of mobile $H_2$ ($D_2$) molecules. We observe the disappearance of hydrogen-induced features from our spectra at 20 ± 5 K, in agreement with Gupta [13], possibly due to capillarity in the tip-surface contact. Finally, we stress that in our experiments we so far did not find any systematic difference in the behaviour of $H_2$ and $D_2$.

We note that the three-dimensional appearance of some of the geometric STHM images (cf. e.g. Fig. 1a) can be ascribed to a slight asymmetry of the physisorption potential on the side of the tip.

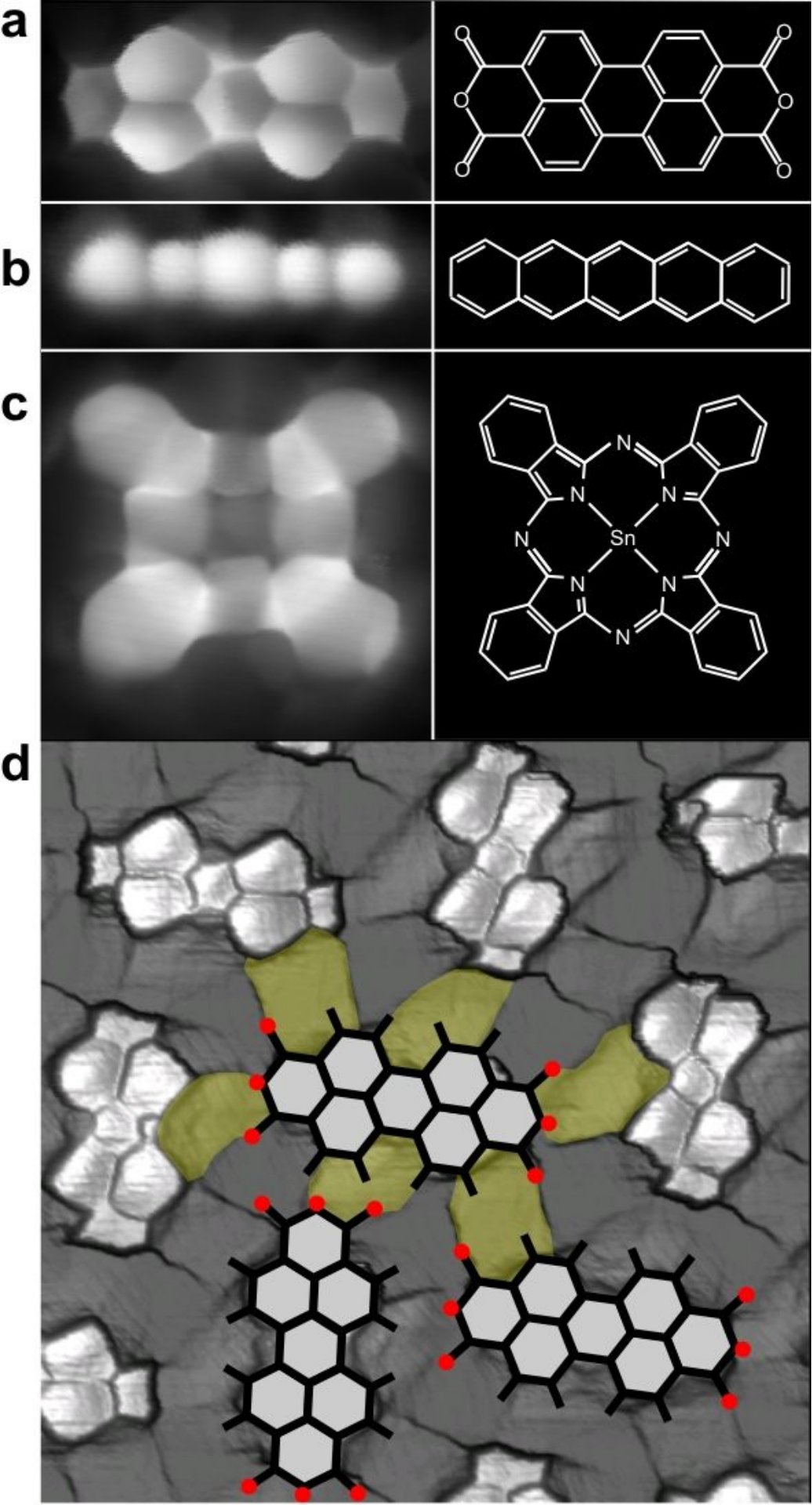

Fig 1, Weiss et al. 2009

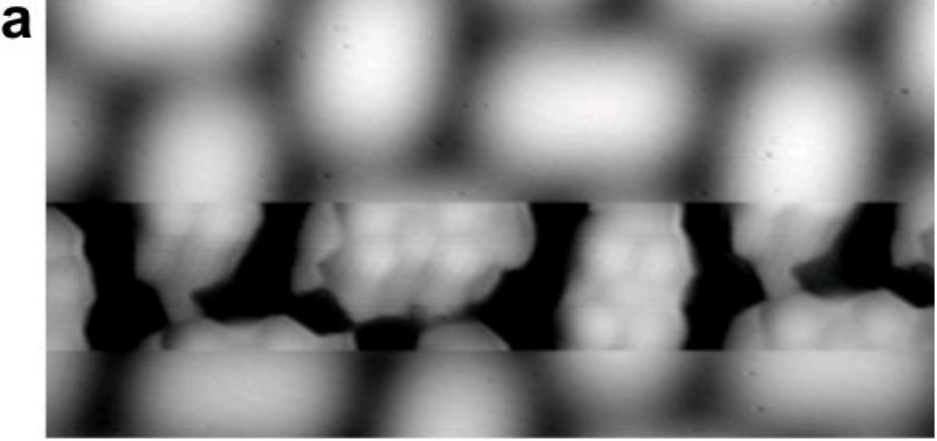
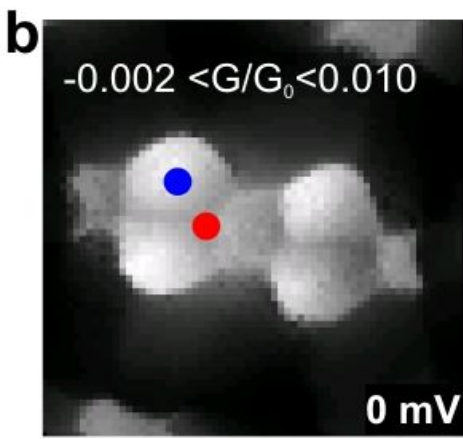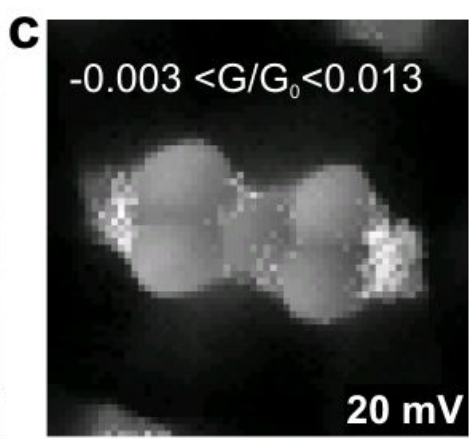
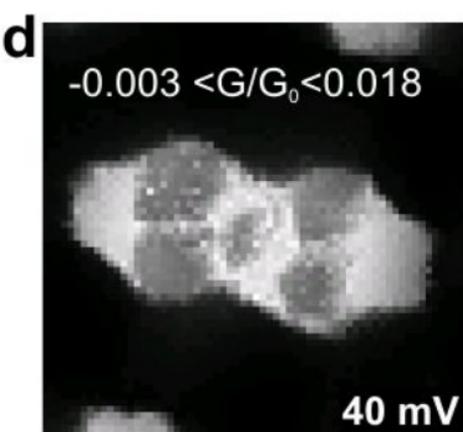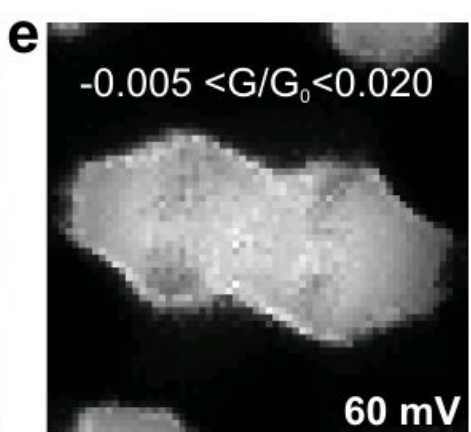
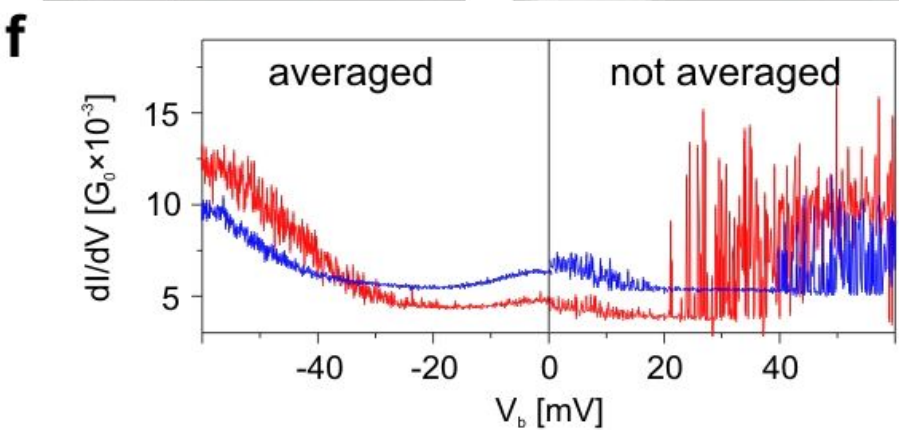

Fig 2, Weiss et al. 2009

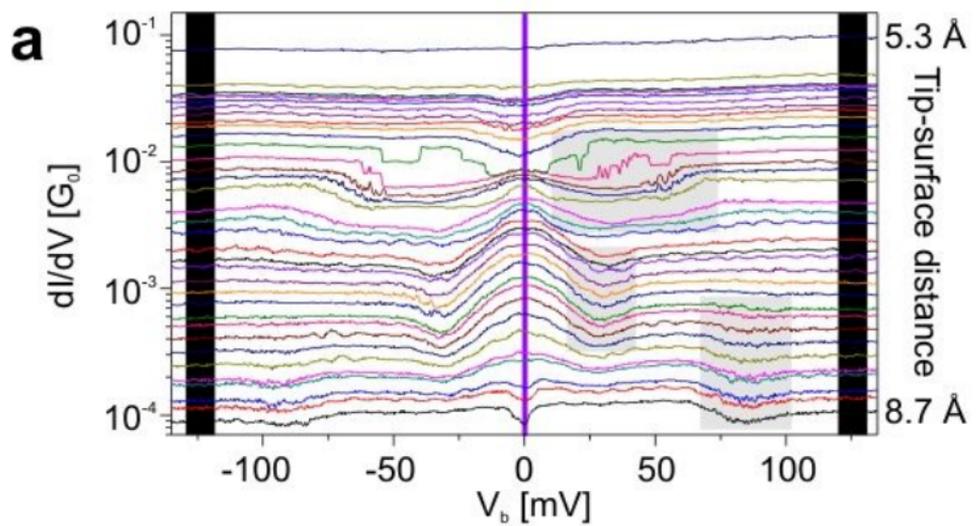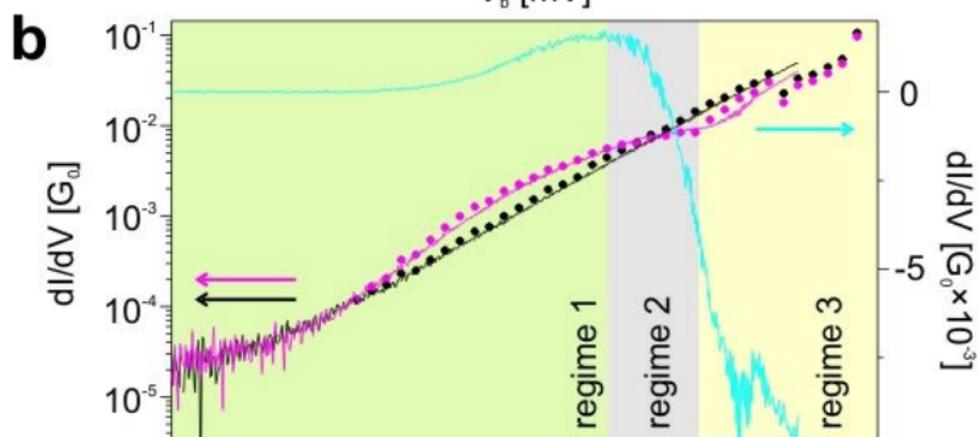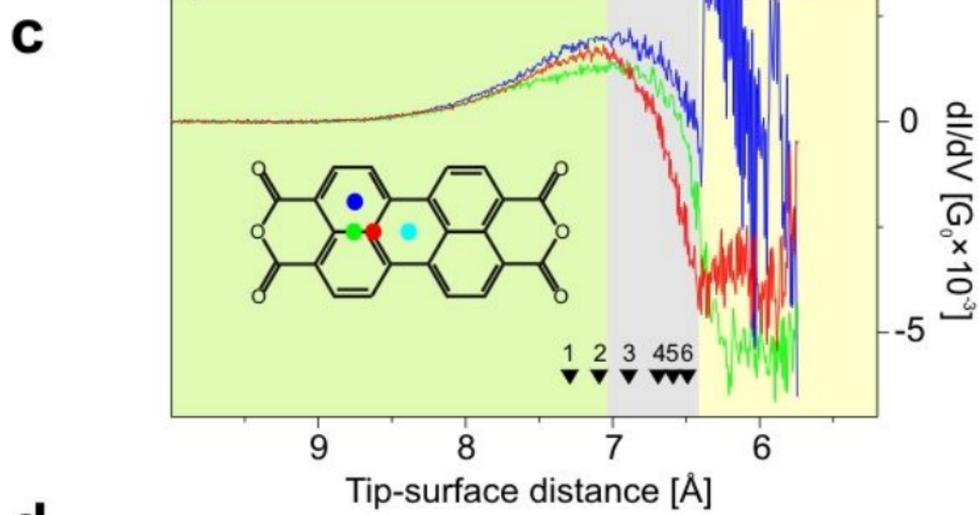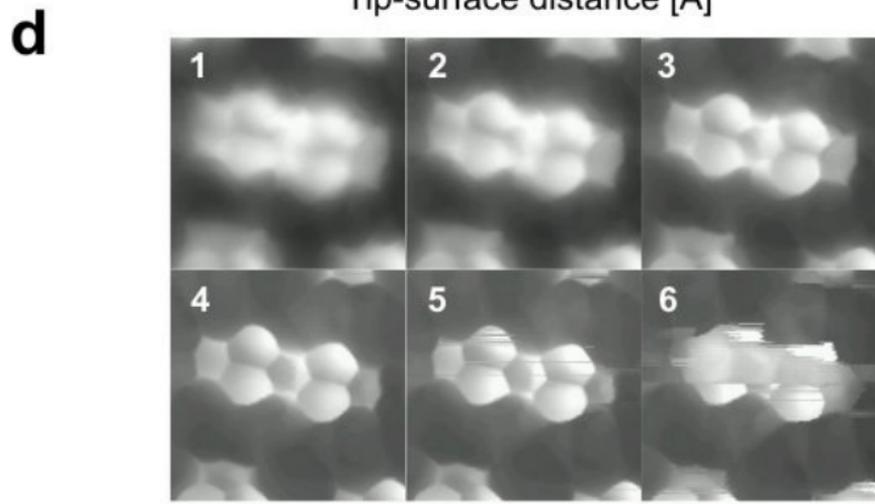

Fig 3, Weiss et al. 2009

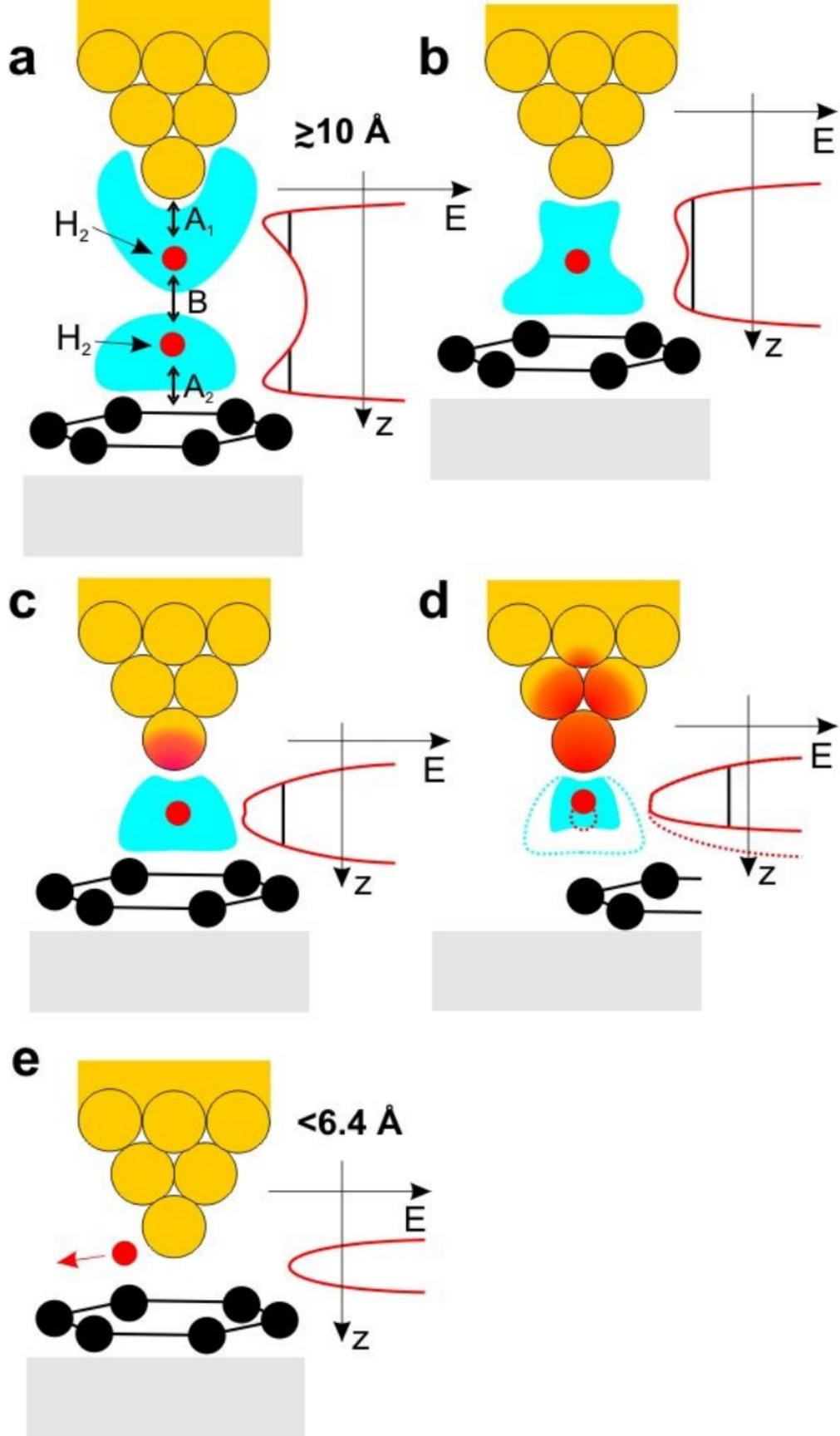

Fig 4, Weiss et al. 2009

**Supplementary Information**

**Figure S1.** STHM image of the PTCDA/Au(111) obtained with $D_2$. The contrast scale of the left panel is adjusted to highlight the intermolecular bond structure. Deterioration of the contrast visible in the bottom right corner of the right panel is due to the sample surface tilt. Imaging parameters: size 5x5 $nm^2$, $V_b$=-10 mV, constant height, current scale left (right) 30 pA (5 pA).

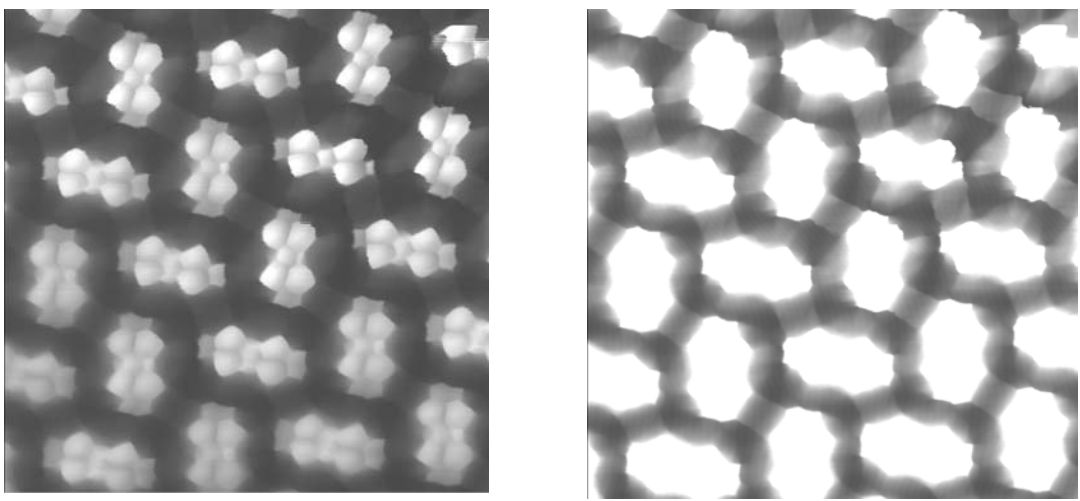

**Figure S2.** Selected dI/dV spectra from Fig. 2a of the main paper shown on a linear scale. Tip-surface distances at which the spectra were measured are indicated. To arrange the spectra on the same scale multiplication factors were applied.

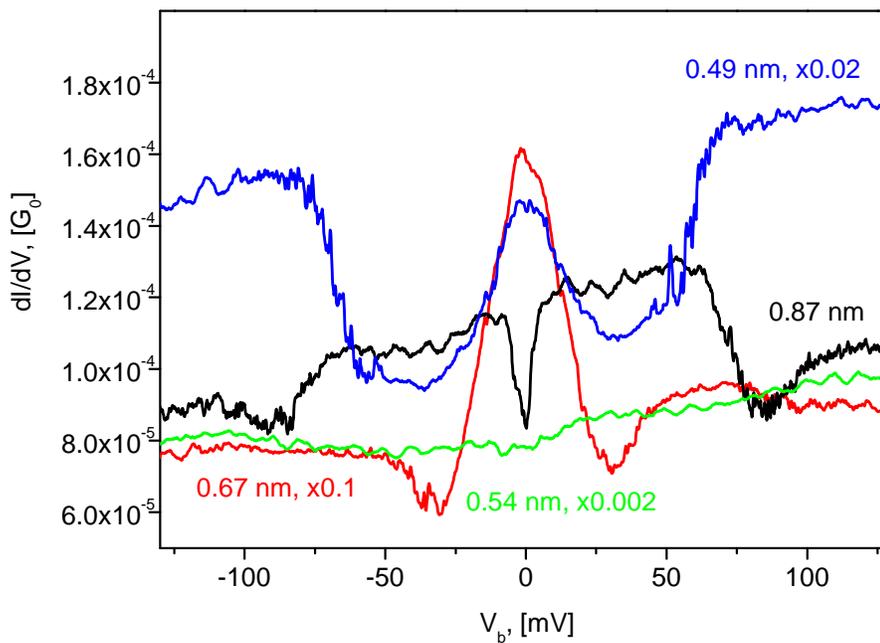

1